\documentclass[prl,twocolumn,showpacs,aps,superscriptaddress,floatfix,letterpaper]{revtex4}
\usepackage{graphicx}  
\usepackage{dcolumn}   
\usepackage{amssymb}   
\usepackage{rotating}  
\usepackage{appendix}
\def\MET{{\mbox{$E\kern-0.57em\raise0.19ex\hbox{/}_{T}$}}}
\def\met{{\mbox{$E\kern-0.57em\raise0.19ex\hbox{/}_{T}$}}}

\def\WH{${\mathit{WH}}\rightarrow \ell\nu b\bar{b}$}

\def\lmet{${\mathit{WH}}\rightarrow \ell\kern-0.45em\raise0.19ex\hbox{/} \nu b\bar{b}$}
\def\ZH{${\mathit{ZH}}\rightarrow \nu\bar{\nu} b\bar{b}$}
\def\ZHll{${\mathit{ZH}}\rightarrow \ell^+ \ell^- b\bar{b}$}

\newcommand{\PYTHIA}     {{\sc pythia}}
\newcommand{\MADGRAPH}     {{\sc madgraph}}
\newcommand{\HDECAY}       {{\sc hdecay}}
\newcommand{\MCFM}       {{\sc mcfm}}
\newcommand{\HERWIG}       {{\sc herwig}}
\newcommand{\MCNLO}       {{\sc mc@nlo}}
\newcommand{\ALPGEN}       {{\sc alpgen}}
\newcommand{\SINGLETOP}       {{\sc singletop}}

\newcommand{\jpzm}{$J^{P}=0^{-}$}
\newcommand{\jptp}{$J^{P}=2^{+}$}

\newcommand{\zhl}{${\mathit{ZH}}\rightarrow \ell\ell b\bar{b}$}

\newcommand{\lvbb}{$\ell\nu b\bar{b}$}
\newcommand{\vvbb}{$\nu\nu b\bar{b}$}

\begin{document}

\hspace{5.2in} \mbox{\begin{tabular}{l} \\
    FERMILAB-PUB-15-029-E \\
\end{tabular}} 



\title{{\vskip 0.5in} Tevatron Constraints on Models of the Higgs Boson with Exotic Spin and Parity Using Decays
to Bottom-Antibottom Quark Pairs}

\affiliation{Institute of Physics, Academia Sinica, Taipei, Taiwan 11529, Republic of China}
\affiliation{Argonne National Laboratory, Argonne, Illinois 60439, USA}
\affiliation{University of Athens, 157 71 Athens, Greece}
\affiliation{Institut de Fisica d'Altes Energies, ICREA, Universitat Autonoma de Barcelona, E-08193, Bellaterra (Barcelona), Spain}
\affiliation{Baylor University, Waco, Texas 76798, USA}
\affiliation{Istituto Nazionale di Fisica Nucleare Bologna, \ensuremath{^{xx}}University of Bologna, I-40127 Bologna, Italy}
\affiliation{University of California, Davis, Davis, California 95616, USA}
\affiliation{University of California, Los Angeles, Los Angeles, California 90024, USA}
\affiliation{Instituto de Fisica de Cantabria, CSIC-University of Cantabria, 39005 Santander, Spain}
\affiliation{Carnegie Mellon University, Pittsburgh, Pennsylvania 15213, USA}
\affiliation{Enrico Fermi Institute, University of Chicago, Chicago, Illinois 60637, USA}
\affiliation{Comenius University, 842 48 Bratislava, Slovakia; Institute of Experimental Physics, 040 01 Kosice, Slovakia}
\affiliation{Joint Institute for Nuclear Research, RU-141980 Dubna, Russia}
\affiliation{Duke University, Durham, North Carolina 27708, USA}
\affiliation{Fermi National Accelerator Laboratory, Batavia, Illinois 60510, USA}
\affiliation{University of Florida, Gainesville, Florida 32611, USA}
\affiliation{Laboratori Nazionali di Frascati, Istituto Nazionale di Fisica Nucleare, I-00044 Frascati, Italy}
\affiliation{University of Geneva, CH-1211 Geneva 4, Switzerland}
\affiliation{Glasgow University, Glasgow G12 8QQ, United Kingdom}
\affiliation{Harvard University, Cambridge, Massachusetts 02138, USA}
\affiliation{Division of High Energy Physics, Department of Physics, University of Helsinki, FIN-00014, Helsinki, Finland; Helsinki Institute of Physics, FIN-00014, Helsinki, Finland}
\affiliation{University of Illinois, Urbana, Illinois 61801, USA}
\affiliation{The Johns Hopkins University, Baltimore, Maryland 21218, USA}
\affiliation{Institut f\"{u}r Experimentelle Kernphysik, Karlsruhe Institute of Technology, D-76131 Karlsruhe, Germany}
\affiliation{Center for High Energy Physics: Kyungpook National University, Daegu 702-701, Korea; Seoul National University, Seoul 151-742, Korea; Sungkyunkwan University, Suwon 440-746, Korea; Korea Institute of Science and Technology Information, Daejeon 305-806, Korea; Chonnam National University, Gwangju 500-757, Korea; Chonbuk National University, Jeonju 561-756, Korea; Ewha Womans University, Seoul, 120-750, Korea}
\affiliation{Ernest Orlando Lawrence Berkeley National Laboratory, Berkeley, California 94720, USA}
\affiliation{University of Liverpool, Liverpool L69 7ZE, United Kingdom}
\affiliation{University College London, London WC1E 6BT, United Kingdom}
\affiliation{Centro de Investigaciones Energeticas Medioambientales y Tecnologicas, E-28040 Madrid, Spain}
\affiliation{Massachusetts Institute of Technology, Cambridge, Massachusetts 02139, USA}
\affiliation{University of Michigan, Ann Arbor, Michigan 48109, USA}
\affiliation{Michigan State University, East Lansing, Michigan 48824, USA}
\affiliation{Institution for Theoretical and Experimental Physics, ITEP, Moscow 117259, Russia}
\affiliation{University of New Mexico, Albuquerque, New Mexico 87131, USA}
\affiliation{The Ohio State University, Columbus, Ohio 43210, USA}
\affiliation{Okayama University, Okayama 700-8530, Japan}
\affiliation{Osaka City University, Osaka 558-8585, Japan}
\affiliation{University of Oxford, Oxford OX1 3RH, United Kingdom}
\affiliation{Istituto Nazionale di Fisica Nucleare, Sezione di Padova, \ensuremath{^{yy}}University of Padova, I-35131 Padova, Italy}
\affiliation{University of Pennsylvania, Philadelphia, Pennsylvania 19104, USA}
\affiliation{Istituto Nazionale di Fisica Nucleare Pisa, \ensuremath{^{zz}}University of Pisa, \ensuremath{^{aaa}}University of Siena, \ensuremath{^{bbb}}Scuola Normale Superiore, I-56127 Pisa, Italy, \ensuremath{^{ccc}}INFN Pavia, I-27100 Pavia, Italy, \ensuremath{^{ddd}}University of Pavia, I-27100 Pavia, Italy}
\affiliation{University of Pittsburgh, Pittsburgh, Pennsylvania 15260, USA}
\affiliation{Purdue University, West Lafayette, Indiana 47907, USA}
\affiliation{University of Rochester, Rochester, New York 14627, USA}
\affiliation{The Rockefeller University, New York, New York 10065, USA}
\affiliation{Istituto Nazionale di Fisica Nucleare, Sezione di Roma 1, \ensuremath{^{eee}}Sapienza Universit\`{a} di Roma, I-00185 Roma, Italy}
\affiliation{Mitchell Institute for Fundamental Physics and Astronomy, Texas A\&M University, College Station, Texas 77843, USA}
\affiliation{Istituto Nazionale di Fisica Nucleare Trieste, \ensuremath{^{fff}}Gruppo Collegato di Udine, \ensuremath{^{ggg}}University of Udine, I-33100 Udine, Italy, \ensuremath{^{hhh}}University of Trieste, I-34127 Trieste, Italy}
\affiliation{University of Tsukuba, Tsukuba, Ibaraki 305, Japan}
\affiliation{Tufts University, Medford, Massachusetts 02155, USA}
\affiliation{University of Virginia, Charlottesville, Virginia 22906, USA}
\affiliation{Waseda University, Tokyo 169, Japan}
\affiliation{Wayne State University, Detroit, Michigan 48201, USA}
\affiliation{University of Wisconsin, Madison, Wisconsin 53706, USA}
\affiliation{Yale University, New Haven, Connecticut 06520, USA}
\affiliation{LAFEX, Centro Brasileiro de Pesquisas F\'{i}sicas, Rio de Janeiro, Brazil}
\affiliation{Universidade do Estado do Rio de Janeiro, Rio de Janeiro, Brazil}
\affiliation{Universidade Federal do ABC, Santo Andr\'{e}, Brazil}
\affiliation{University of Science and Technology of China, Hefei, People's Republic of China}
\affiliation{Universidad de los Andes, Bogot\'{a}, Colombia}
\affiliation{Charles University, Faculty of Mathematics and Physics, Center for Particle Physics, Prague, Czech Republic}
\affiliation{Czech Technical University in Prague, Prague, Czech Republic}
\affiliation{Institute of Physics, Academy of Sciences of the Czech Republic, Prague, Czech Republic}
\affiliation{Universidad San Francisco de Quito, Quito, Ecuador}
\affiliation{LPC, Universit\'{e} Blaise Pascal, CNRS/IN2P3, Clermont, France}
\affiliation{LPSC, Universit\'{e} Joseph Fourier Grenoble 1, CNRS/IN2P3, Institut National Polytechnique de Grenoble, Grenoble, France}
\affiliation{CPPM, Aix-Marseille Universit\'{e}, CNRS/IN2P3, Marseille, France}
\affiliation{LAL, Universit\'{e} Paris-Sud, CNRS/IN2P3, Orsay, France}
\affiliation{LPNHE, Universit\'{e}s Paris VI and VII, CNRS/IN2P3, Paris, France}
\affiliation{CEA, Irfu, SPP, Saclay, France}
\affiliation{IPHC, Universit\'{e} de Strasbourg, CNRS/IN2P3, Strasbourg, France}
\affiliation{IPNL, Universit\'{e} Lyon 1, CNRS/IN2P3, Villeurbanne, France and Universit\'{e} de Lyon, Lyon, France}
\affiliation{III. Physikalisches Institut A, RWTH Aachen University, Aachen, Germany}
\affiliation{Physikalisches Institut, Universit\"{a}t Freiburg, Freiburg, Germany}
\affiliation{II. Physikalisches Institut, Georg-August-Universit\"{a}t G\"{o}ttingen, G\"{o}ttingen, Germany}
\affiliation{Institut f\"{u}r Physik, Universit\"{a}t Mainz, Mainz, Germany}
\affiliation{Ludwig-Maximilians-Universit\"{a}t M\"{u}nchen, M\"{u}nchen, Germany}
\affiliation{Panjab University, Chandigarh, India}
\affiliation{Delhi University, Delhi, India}
\affiliation{Tata Institute of Fundamental Research, Mumbai, India}
\affiliation{University College Dublin, Dublin, Ireland}
\affiliation{Korea Detector Laboratory, Korea University, Seoul, Korea}
\affiliation{CINVESTAV, Mexico City, Mexico}
\affiliation{Nikhef, Science Park, Amsterdam, the Netherlands}
\affiliation{Radboud University Nijmegen, Nijmegen, the Netherlands}
\affiliation{Joint Institute for Nuclear Research, RU-141980 Dubna, Russia}
\affiliation{Institution for Theoretical and Experimental Physics, ITEP, Moscow 117259, Russia}
\affiliation{Moscow State University, Moscow, Russia}
\affiliation{Institute for High Energy Physics, Protvino, Russia}
\affiliation{Petersburg Nuclear Physics Institute, St. Petersburg, Russia}
\affiliation{Instituci\'{o} Catalana de Recerca i Estudis Avan\c{c}ats (ICREA) and Institut de F\'{i}sica d'Altes Energies (IFAE), Barcelona, Spain}
\affiliation{Uppsala University, Uppsala, Sweden}
\affiliation{Taras Shevchenko National University of Kyiv, Kiev, Ukraine}
\affiliation{Lancaster University, Lancaster LA1 4YB, United Kingdom}
\affiliation{Imperial College London, London SW7 2AZ, United Kingdom}
\affiliation{The University of Manchester, Manchester M13 9PL, United Kingdom}
\affiliation{University of Arizona, Tucson, Arizona 85721, USA}
\affiliation{University of California Riverside, Riverside, California 92521, USA}
\affiliation{Florida State University, Tallahassee, Florida 32306, USA}
\affiliation{Fermi National Accelerator Laboratory, Batavia, Illinois 60510, USA}
\affiliation{University of Illinois at Chicago, Chicago, Illinois 60607, USA}
\affiliation{Northern Illinois University, DeKalb, Illinois 60115, USA}
\affiliation{Northwestern University, Evanston, Illinois 60208, USA}
\affiliation{Indiana University, Bloomington, Indiana 47405, USA}
\affiliation{Purdue University Calumet, Hammond, Indiana 46323, USA}
\affiliation{University of Notre Dame, Notre Dame, Indiana 46556, USA}
\affiliation{Iowa State University, Ames, Iowa 50011, USA}
\affiliation{University of Kansas, Lawrence, Kansas 66045, USA}
\affiliation{Louisiana Tech University, Ruston, Louisiana 71272, USA}
\affiliation{Northeastern University, Boston, Massachusetts 02115, USA}
\affiliation{University of Michigan, Ann Arbor, Michigan 48109, USA}
\affiliation{Michigan State University, East Lansing, Michigan 48824, USA}
\affiliation{University of Mississippi, University, Mississippi 38677, USA}
\affiliation{University of Nebraska, Lincoln, Nebraska 68588, USA}
\affiliation{Rutgers University, Piscataway, New Jersey 08855, USA}
\affiliation{Princeton University, Princeton, New Jersey 08544, USA}
\affiliation{State University of New York, Buffalo, New York 14260, USA}
\affiliation{University of Rochester, Rochester, New York 14627, USA}
\affiliation{State University of New York, Stony Brook, New York 11794, USA}
\affiliation{Brookhaven National Laboratory, Upton, New York 11973, USA}
\affiliation{Langston University, Langston, Oklahoma 73050, USA}
\affiliation{University of Oklahoma, Norman, Oklahoma 73019, USA}
\affiliation{Oklahoma State University, Stillwater, Oklahoma 74078, USA}
\affiliation{Brown University, Providence, Rhode Island 02912, USA}
\affiliation{University of Texas, Arlington, Texas 76019, USA}
\affiliation{Southern Methodist University, Dallas, Texas 75275, USA}
\affiliation{Rice University, Houston, Texas 77005, USA}
\affiliation{University of Virginia, Charlottesville, Virginia 22904, USA}
\affiliation{University of Washington, Seattle, Washington 98195, USA}

\author{T.~Aaltonen \ensuremath{^{\dagger}}}
\affiliation{Division of High Energy Physics, Department of Physics, University of Helsinki, FIN-00014, Helsinki, Finland; Helsinki Institute of Physics, FIN-00014, Helsinki, Finland}
\author{V.M.~Abazov \ensuremath{^{\ddagger}}}
\affiliation{Joint Institute for Nuclear Research, RU-141980 Dubna, Russia}
\author{B.~Abbott \ensuremath{^{\ddagger}}}
\affiliation{University of Oklahoma, Norman, Oklahoma 73019, USA}
\author{B.S.~Acharya \ensuremath{^{\ddagger}}}
\affiliation{Tata Institute of Fundamental Research, Mumbai, India}
\author{M.~Adams \ensuremath{^{\ddagger}}}
\affiliation{University of Illinois at Chicago, Chicago, Illinois 60607, USA}
\author{T.~Adams \ensuremath{^{\ddagger}}}
\affiliation{Florida State University, Tallahassee, Florida 32306, USA}
\author{J.P.~Agnew \ensuremath{^{\ddagger}}}
\affiliation{The University of Manchester, Manchester M13 9PL, United Kingdom}
\author{G.D.~Alexeev \ensuremath{^{\ddagger}}}
\affiliation{Joint Institute for Nuclear Research, RU-141980 Dubna, Russia}
\author{G.~Alkhazov \ensuremath{^{\ddagger}}}
\affiliation{Petersburg Nuclear Physics Institute, St. Petersburg, Russia}
\author{A.~Alton \ensuremath{^{\ddagger}}\ensuremath{^{jj}}}
\affiliation{University of Michigan, Ann Arbor, Michigan 48109, USA}
% ...

\collaboration{CDF Collaboration}
\altaffiliation[With visitors from]{
\ensuremath{^{a}}University of British Columbia, Vancouver, BC V6T 1Z1, Canada,
\ensuremath{^{b}}Istituto Nazionale di Fisica Nucleare, Sezione di Cagliari, 09042 Monserrato (Cagliari), Italy,
\ensuremath{^{c}}University of California Irvine, Irvine, CA 92697, USA,
\ensuremath{^{d}}Institute of Physics, Academy of Sciences of the Czech Republic, 182 21, Czech Republic,
\ensuremath{^{e}}CERN, CH-1211 Geneva, Switzerland,
\ensuremath{^{f}}Cornell University, Ithaca, NY 14853, USA,
\ensuremath{^{g}}University of Cyprus, Nicosia CY-1678, Cyprus,
\ensuremath{^{h}}Office of Science, U.S. Department of Energy, Washington, DC 20585, USA,
\ensuremath{^{i}}University College Dublin, Dublin 4, Ireland,
\ensuremath{^{j}}ETH, 8092 Z\"{u}rich, Switzerland,
\ensuremath{^{k}}University of Fukui, Fukui City, Fukui Prefecture, Japan 910-0017,
\ensuremath{^{l}}Universidad Iberoamericana, Lomas de Santa Fe, M\'{e}xico, C.P. 01219, Distrito Federal,
\ensuremath{^{m}}University of Iowa, Iowa City, IA 52242, USA,
\ensuremath{^{n}}Kinki University, Higashi-Osaka City, Japan 577-8502,
\ensuremath{^{o}}Kansas State University, Manhattan, KS 66506, USA,
\ensuremath{^{p}}Brookhaven National Laboratory, Upton, NY 11973, USA,
\ensuremath{^{q}}Istituto Nazionale di Fisica Nucleare, Sezione di Lecce, Via Arnesano, I-73100 Lecce, Italy,
\ensuremath{^{r}}Queen Mary, University of London, London, E1 4NS, United Kingdom,
\ensuremath{^{s}}University of Melbourne, Victoria 3010, Australia,
\ensuremath{^{t}}Muons, Inc., Batavia, IL 60510, USA,
\ensuremath{^{u}}Nagasaki Institute of Applied Science, Nagasaki 851-0193, Japan,
\ensuremath{^{v}}National Research Nuclear University, Moscow 115409, Russia,
\ensuremath{^{w}}Northwestern University, Evanston, IL 60208, USA,
\ensuremath{^{x}}University of Notre Dame, Notre Dame, IN 46556, USA,
\ensuremath{^{y}}Universidad de Oviedo, E-33007 Oviedo, Spain,
\ensuremath{^{z}}CNRS-IN2P3, Paris, F-75205 France,
\ensuremath{^{aa}}Universidad Tecnica Federico Santa Maria, 110v Valparaiso, Chile,
\ensuremath{^{bb}}The University of Jordan, Amman 11942, Jordan,
\ensuremath{^{cc}}Universite catholique de Louvain, 1348 Louvain-La-Neuve, Belgium,
\ensuremath{^{dd}}University of Z\"{u}rich, 8006 Z\"{u}rich, Switzerland,
\ensuremath{^{ee}}Massachusetts General Hospital, Boston, MA 02114 USA,
\ensuremath{^{ff}}Harvard Medical School, Boston, MA 02114 USA,
\ensuremath{^{gg}}Hampton University, Hampton, VA 23668, USA,
\ensuremath{^{hh}}Los Alamos National Laboratory, Los Alamos, NM 87544, USA,
\ensuremath{^{ii}}Universit\`{a} degli Studi di Napoli Federico I, I-80138 Napoli, Italy
}
\noaffiliation
\collaboration{D0 Collaboration}
\altaffiliation[With visitors from]{
\ensuremath{^{jj}}Augustana College, Sioux Falls, SD, USA,
\ensuremath{^{kk}}The University of Liverpool, Liverpool, UK,
\ensuremath{^{ll}}DESY, Hamburg, Germany,
\ensuremath{^{mm}}Universidad Michoacana de San Nicolas de Hidalgo, Morelia, Mexico,
\ensuremath{^{nn}}SLAC, Menlo Park, CA, USA,
\ensuremath{^{oo}}University College London, London, UK,
\ensuremath{^{pp}}Centro de Investigacion en Computacion - IPN, Mexico City, Mexico,
\ensuremath{^{qq}}Universidade Estadual Paulista, S\~{a}o Paulo, Brazil,
\ensuremath{^{rr}}Karlsruher Institut f\"{u}r Technologie (KIT) - Steinbuch Centre for Computing (SCC),
\ensuremath{^{ss}}Office of Science, U.S. Department of Energy, Washington, D.C. 20585, USA,
\ensuremath{^{tt}}American Association for the Advancement of Science, Washington, D.C. 20005, USA,
\ensuremath{^{uu}}National Academy of Science of Ukraine (NASU) - Kiev Institute for Nuclear Research (KINR),
\ensuremath{^{vv}}University of Maryland, College Park, MD 20742,
\ensuremath{^{ww}}European Organization for Nuclear Research (CERN)
}
\noaffiliation

\date{February 3, 2015}

\begin{abstract}
Combined constraints from the CDF and D0 Collaborations on models
of the Higgs boson with exotic spin $J$ and parity $P$ are presented and compared with results obtained
assuming the standard model value $J^P=0^+$.
Both collaborations analyzed approximately 10~fb$^{-1}$ of proton-antiproton collisions 
with a center-of-mass energy of 1.96~TeV collected
at the Fermilab Tevatron.  
Two models predicting exotic Higgs bosons with $J^P=0^-$ and $J^P=2^+$ are tested. 
The kinematic properties of exotic Higgs boson production in association with a vector
boson differ from those predicted for the standard model Higgs boson.
Upper limits at the 95\% credibility level on the production rates of the exotic
Higgs bosons, expressed as fractions of the standard model Higgs boson production rate,
are set at 0.36 for both the $J^P=0^-$ hypothesis and the
$J^P=2^+$ hypothesis.
If the production rate times the
branching ratio to a bottom-antibottom pair is
the same as that predicted for the standard model Higgs boson, then the exotic
bosons are excluded with significances of 5.0 standard deviations and 4.9 standard deviations for
the $J^P=0^-$ and $J^P=2^+$ hypotheses, respectively.
\end{abstract}


\maketitle

The Higgs boson discovered by the
ATLAS~\cite{atlasdiscovery} and CMS~\cite{cmsdiscovery} Collaborations
in 2012 using data produced in proton-proton collisions at the Large Hadron Collider (LHC) at CERN
allows many stringent tests of the electroweak symmetry breaking in the standard model (SM) and
extensions to the SM to be performed.  
To date, measurements of the Higgs boson's 
mass and width~\cite{cmsmass,atlasmass,atlaswidth,cmswidth}, its couplings to other 
particles~\cite{atlascouplings,atlascouplings2,atlascouplings3,cmsmass,cmszz4l,cmsdiphotons}, 
and its spin and parity quantum numbers 
$J$ and $P$~\cite{atlasjp1,atlasjp2,cmsjp,cmsjp2,cmszz4l,cmsdiphotons,cmsjp14} are consistent with 
the expectations for the SM Higgs boson.  
The CDF and D0 Collaborations at the Fermilab Tevatron observed
a 3.0~standard deviation (s.d.) excess of events consistent with a Higgs boson signal, largely driven 
by those channels sensitive to the decay of the 
Higgs boson to bottom quarks ($H\rightarrow b{\bar{b}}$)~\cite{tevbbevidence,tevstudies}.
The Tevatron data are also consistent with the predictions
for the properties of the SM Higgs boson~\cite{cdfsmcomb,d0smcomb,tevstudies,cdfjp,d0jp}.

Ref.~\cite{Ellis:2012xd} proposed
to use the Tevatron data to test models for the 
Higgs boson with exotic spin and parity, using events in which
the exotic Higgs boson $X$ is produced in association 
with a $W$ or a $Z$ boson and decays to a bottom-antibottom quark pair, $X\rightarrow b\bar{b}$.
This proposal used two of the spin and
parity models in Ref.~\cite{Miller:2001bi}, one with a pseudoscalar $J^P=0^-$ state
and the other with a graviton-like $J^P=2^+$ state.  For the SM Higgs boson, which has $J^P=0^+$,
the differential production rate near threshold is linear in 
$\beta$, where $\beta = 2p/\sqrt{\hat{s}}$, $p$ is the momentum of the $X$ boson in the
{\it VX} ($V$ = $W$ or $Z$) reference frame, and  $\sqrt{\hat{s}}$ is the total energy of the {\it VX} system 
in its rest frame. For the pseudoscalar model, the dependence is proportional to $\beta^3$. For the graviton-like model, 
the dependence is proportional to $\beta^5$; however, not all $J^P=2^+$ models share
this $\beta^5$ factor~\cite{Ellis:2012xd}. 
These powers of $\beta$ alter the kinematic distributions of the observable decay products of the vector boson 
and the Higgs-like boson $X$, most notably the invariant mass of the {\it VX} system, which has a higher average
value in the $J^P=0^-$ hypothesis than in the SM $0^+$ case, and higher still in the $J^P=2^+$ hypothesis.  These models
predict neither the production rates nor the decay branching fractions of the $X$ particles.

The ATLAS and CMS Collaborations recently reported strong evidence for Higgs boson decays to 
fermions~\cite{Chatrchyan:2014vua,cmstt,atlastau,atlastt,cmsbb,atlasbb}, with sensitivity
dominated by the $H\rightarrow\tau^+\tau^-$ decay mode, though they have not yet performed
spin and parity tests using fermionic decays.
The particle decaying fermionically for which the Tevatron also found evidence
might not be the same as the particle discovered through its bosonic decays at the LHC.
Tests of the spin and parity~\cite{Ellis:2012xd}
with Tevatron data therefore provide unique information on the identity 
and properties of the new particle or particles.
The CDF and D0 Collaborations have re-optimized their SM Higgs boson 
searches to test the exotic Higgs boson models in the 
${\mathit{WH}}\rightarrow\ell\nu b{\bar{b}}$~\cite{cdfWHpub,dzwhl}, 
${\mathit{ZH}}\rightarrow\ell^+\ell^- b{\bar{b}}$~\cite{cdfZHllpub,dzzhll}, and 
${\mathit{WH}}+{\mathit{ZH}}\rightarrow\MET b{\bar{b}}$~\cite{cdfmetbbpub,dzzhv2} channels, 
where $\ell = e$ or $\mu$ and $\MET$ is the missing transverse energy~\cite{coord}.
In this letter we report a combination of the CDF~\cite{cdfjp} and D0~\cite{d0jp} 
studies of the $J^P$ assignments of the state $X$, with mass $m_X = 125$~GeV/$c^2$, 
in the $X\rightarrow b{\bar{b}}$ decay.

The CDF and D0 detectors are multipurpose solenoidal spectrometers surrounded by 
hermetic calorimeters and muon detectors designed to study the products
of 1.96 TeV proton-antiproton ($p{\bar{p}}$) collisions~\cite{cdfdetector,d0detector}.
All searches combined here use the complete Tevatron data sample,
which, after data quality requirements, corresponds to 9.45 -- 9.7~fb$^{-1}$ of integrated luminosity,
depending on the experiment and the search channel.

Standard model Higgs boson signal events are simulated using the leading-order (LO) calculation from 
\PYTHIA~\cite{pythia},
with CTEQ5L (CDF) and CTEQ6L1 (D0)~\cite{cteq} parton distribution functions (PDFs).
The  \jpzm\ and \jptp\ signal samples are generated using
\MADGRAPH~5 version $1.4.8.4$~\cite{madgraph}, with modifications provided by the authors of Ref.~\cite{Ellis:2012xd}.
Subsequent particle showering is modeled by \PYTHIA.
We normalize the SM Higgs boson rate predictions to the highest-order calculations
available.  The {\it WH} and {\it ZH} cross sections are
calculated at next-to-next-to-leading-order (NNLO) precision in the strong interaction, and
next-to-leading-order (NLO) precision in the 
electroweak corrections~\cite{vhtheory,vhtheory2,vhtheory3,vhtheory4}.  We use the branching fractions
for Higgs boson decay from Ref.~\cite{lhcdifferential}.
These rely on calculations using \HDECAY~\cite{hdecay}
and {\sc prophecy4f}~\cite{prophecy4f}.

The predictions of the dominant background rates and kinematic distributions
are treated in the following way.
Diboson ({\it{WW}}, {\it{WZ}},  and {\it{ZZ}}) Monte Carlo (MC) samples are normalized using 
the NLO calculations from \MCFM~\cite{mcfm}.  For $t{\bar{t}}$, we use a
production cross section of $7.04\pm 0.70$~pb~\cite{mochuwer}, which is
based on a top quark mass of 173~GeV/$c^2$~\cite{topmass} and MSTW 2008 
NNLO PDFs~\cite{mstw2008}.  The single top quark production cross section is 
assumed to be $3.15\pm 0.31$~pb~\cite{kidonakis_st}. For details of the 
generators used, see Ref.~\cite{mcbla}.  Data-driven methods 
are used to normalize the {\it V} plus light-flavor and heavy-flavor 
jet backgrounds~\cite{hfjet} using {\it V} data events containing no
$b$-tagged jets~\cite{btagdef}, which have negligible signal content~\cite{cdfHbb2012,dzHbb2012}.
The MC modeling of the kinematic distributions of the background predictions is described in 
Refs.~\cite{cdfWHpub,dzwhl,cdfZHllpub,dzzhll,cdfmetbbpub,dzzhv2}.

The event selections are similar (CDF), or identical (D0), to 
those used in their SM counterparts~\cite{cdfWHpub,dzwhl,cdfZHllpub,dzzhll,cdfmetbbpub,dzzhv2}.
For the \WH\ analyses, events are selected with one identified lepton ($e$ or $\mu$), jets,
and large \met.
For the CDF \WH\ analysis, only events with two jets are used.    Events are classified
into separate categories based on the quality of the
identified lepton.  Separate categories are used for events with a
high-quality muon or central electron candidate, an isolated track, or
a forward electron candidate.  Within the lepton
categories, five exclusive $b$-tag categories, comprising 
two single-tag and three double-tag categories, are formed.
The multivariate $b$-tagger used by CDF~\cite{Freeman:2012uf} was trained on SM Higgs boson signal 
MC events.  Few of these events contained
jets with with transverse energy $E_T>200$~GeV and thus the tagger
does not perform well for such jets. Hence, only jets with $E_T<200$~GeV are considered. For the D0 \WH\ analysis, events are selected with two or three jets.
The data are split by lepton flavor and jet
multiplicity (two or three jet subchannels), and by the output of the $b$-tagging algorithm applied to all 
selected jets in the event.  This channel, along with the other two D0 channels, uses a multivariate 
$b$-tagging algorithm~\cite{Abazov:2010ab,dzbtag2}.
Four exclusive $b$-tag categories, one single-tag and three double-tag, are formed.
In the SM Higgs boson search, boosted decision trees are used as 
the final discriminating variables; here they are used to further subdivide the
selected data sample into high- and low-purity categories.

The \ZHll\ analyses require two isolated leptons and at least two jets.  
The CDF analysis separates events into one single- and three double-$b$-tag samples and uses 
neural networks to select loose dielectron and dimuon candidates.
The jet energies are corrected for \met\ using a neural 
network~\cite{cdfllbb2010}. The CDF analysis uses a multistep discriminant based on
neural networks, where two discriminant functions are used to define 
three separate regions of the final discriminant function. 
The D0 \ZHll\ analysis separates events into non-overlapping
samples of events with either a single or double $b$-tag. To increase the 
signal acceptance, the selection criteria for one
of the leptons are loosened to include isolated tracks not reconstructed in the
muon detector 
and electron candidates from the intercryostat
region of the D0 detector. 
Combined with the dielectron
and dimuon 
categories, these provide four independent lepton
subchannels. A kinematic fit is used to optimize 
reconstruction. Random forests (RF) of decision trees~\cite{forest,tmva} are used to provide the final variables
in the SM Higgs boson search.  The first RF is designed to 
discriminate against $t\bar{t}$ events and divides events into $t\bar{t}$-enriched and $t\bar{t}$-depleted
single-tag and double-tag regions. Only events in the $t\bar{t}$-depleted regions are considered in this study. 
These regions contain approximately 94\% of the SM signal. 

For the \ZH\ analyses, the selections used by CDF and D0 are similar to the ${\mathit{WH}}$ selections,
except that all events with isolated leptons are rejected and more stringent
techniques are applied to reject the multijet background.
In a sizable fraction
of \WH\ signal events, the lepton is undetected.  Such events often are selected
in the \ZH\ samples, so these analyses are also referred to as $VH\rightarrow \met b \bar{b}$.  The CDF analysis uses three
non-overlapping $b$-tag categories (two double- and one single-tag) and two
jet categories (two- or three-jet events), giving a total of six subchannels. 
In the D0 analysis, exactly two jets are required and two exclusive double-tag categories are defined using the sum of 
the $b$-tagging outputs for each of the two selected jets. 

Both CDF and D0 have a 50\% larger acceptance for the $J^P=0^-$ and $J^P=2^+$ signals in the \ZH\ analyses 
compared with the SM Higgs boson signal, largely due
to the fact that the exotic signal events are more likely to pass the trigger thresholds
for \met, a consequence of the larger average $VX$ invariant masses.  
The other two channels, \WH\ and \ZHll, do not benefit as much
from the additional $\met$ in these events, as they rely on the lepton triggers,
which are more efficient than the $\met$ triggers in the relevant kinematic regions.

Unlike their SM counterparts, these analyses 
are optimized to distinguish the $J^P=0^-$ and the $J^P=2^+$ hypotheses from the SM $0^+$ hypothesis.
The exotic particles are considered either in addition to, or replacing, the SM Higgs boson.  
A mixture of all three states is
not considered.   
 
The CDF multivariate analysis (MVA) discriminants were newly trained to separate the exotic Higgs boson signals
from the SM backgrounds~\cite{cdfjp}. 
In the ${\mathit{WH}}\rightarrow\ell\nu b{\bar{b}}$ and $VH\rightarrow\MET b{\bar{b}}$ channels,
events classified as background-like by the new discriminants are then classified
according to the SM-optimized MVA discriminants in order to improve 
the performance of tests between the SM and exotic hypotheses.

 Depending on the channel, D0 uses either 
the reconstructed dijet mass or the MVA used in the SM Higgs boson search to separate events into high- and low-purity samples.
The mass of the {\it VX} system is then used to discriminate between
the exotic and SM hypotheses~\cite{d0jp}. For the 
\zhl\ analysis the invariant mass of the two leptons and the two highest $p_T$ jets is used. For the \lvbb\ and \vvbb\ final states the transverse mass $M_T$ is used, where
$M_{T}^{2}=(E_{T}^{V}+E_{T}^{X})^{2}-(\vec{p}_{T}^{V}+\vec{p}_{T}^{X})^{2}$ and
the transverse momenta of the $Z$ and $W$ bosons are taken to be $\vec{p}_{T}^{Z}=\vec{\not\!\!E}_{T}$ and $\vec{p}_{T}^{W}=\vec{\not\!\!E}_{T}+\vec{p}_{T}^{\ell}$, respectively.

\begin{figure}
\includegraphics[width=0.8\columnwidth]{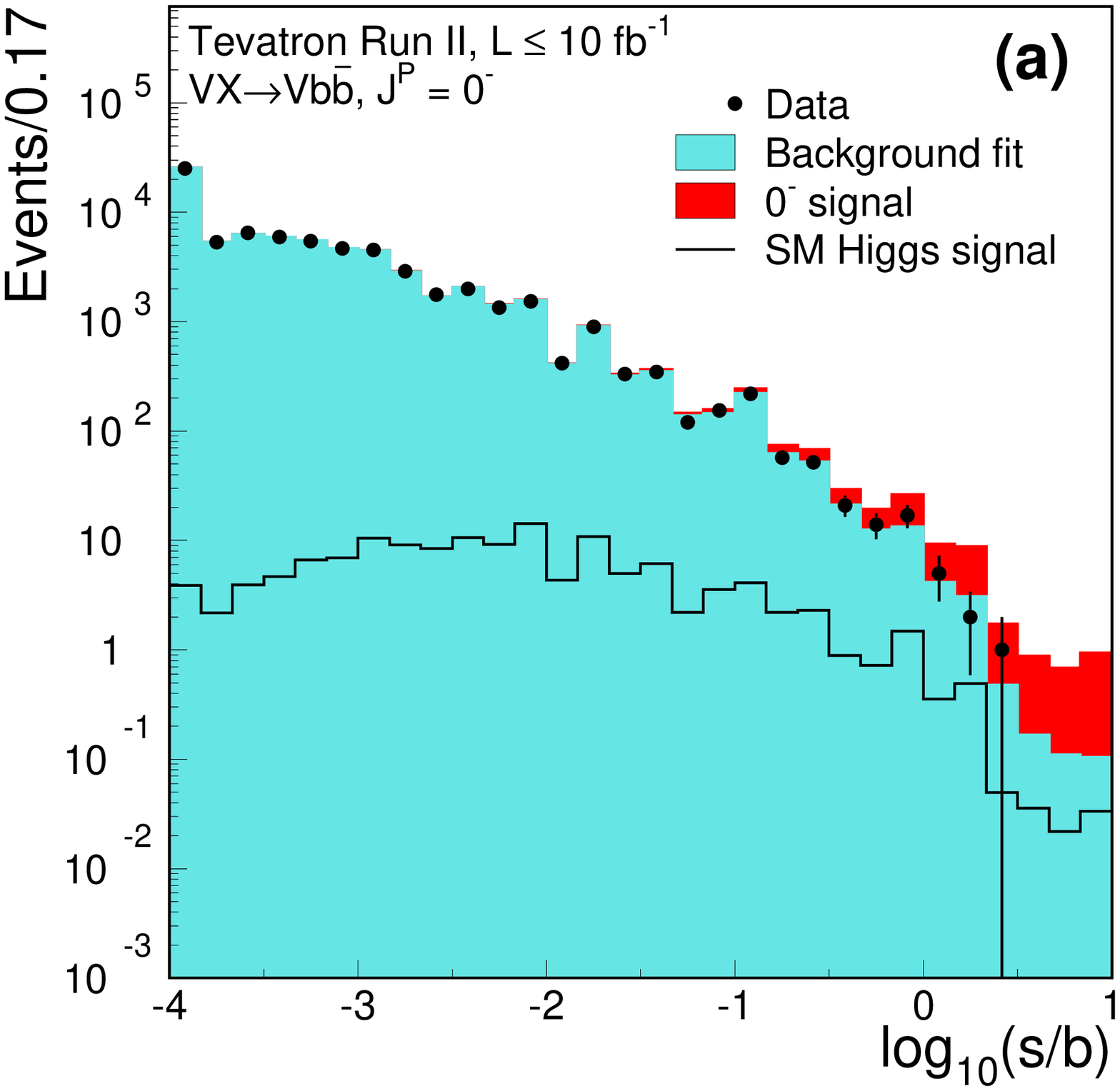}
\includegraphics[width=0.8\columnwidth]{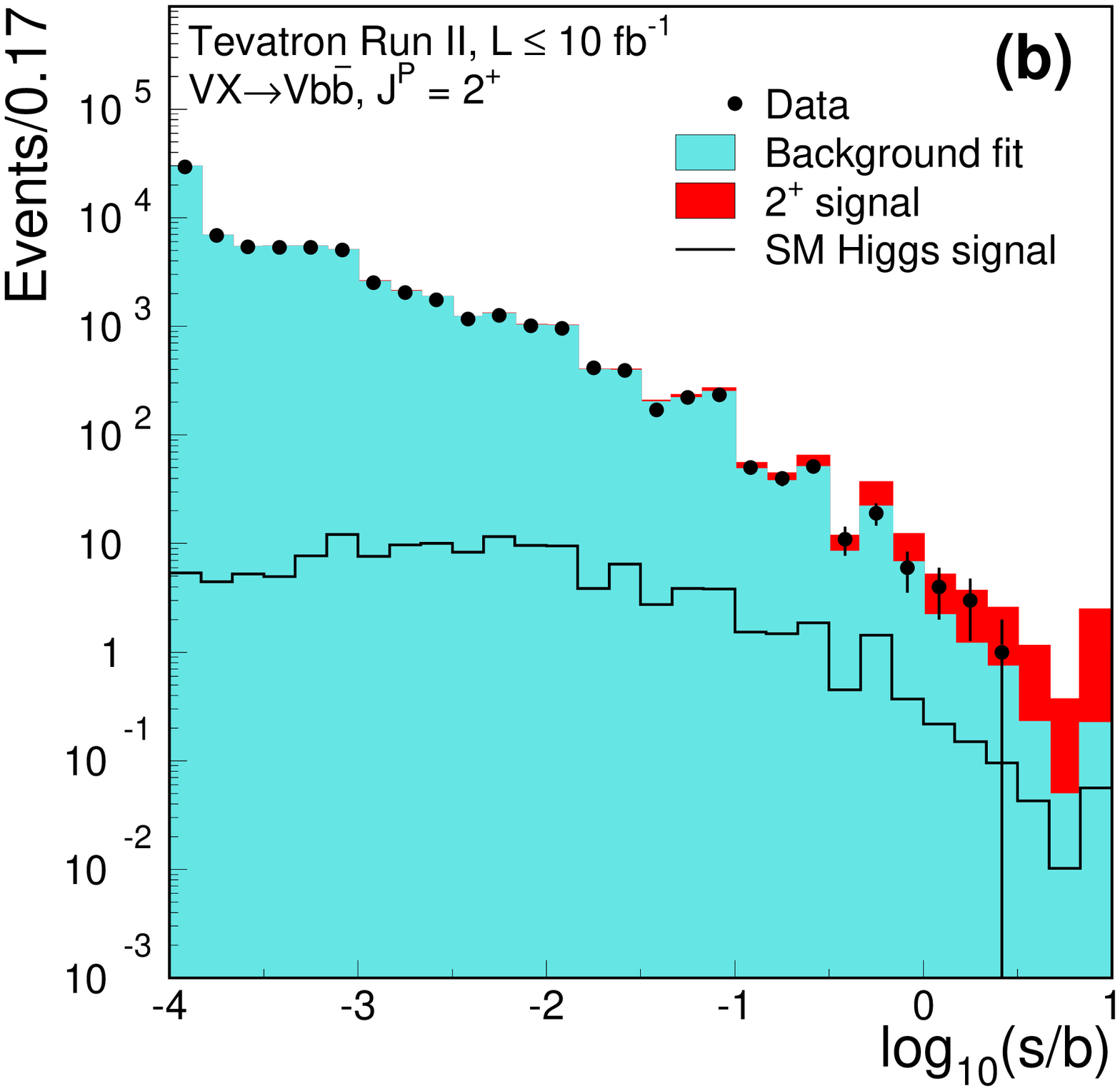}
\caption{ \label{fig:lnsb} (color online).  Distribution of $\log_{10}(s/b)$ for CDF and D0 data 
from all contributing search channels, for (a) the
$J^P=0^-$ search and (b) the $J^P=2^+$ search.  
The data are shown with points, and the expected exotic signals are shown with $\mu_{\rm{exotic}}=1$ 
stacked on top of the  fitted backgrounds. The solid
lines denote the predictions for the SM Higgs boson, and are not stacked. 
Underflows and overflows are collected into the leftmost and rightmost bins, respectively.}
\end{figure}

The number of contributing channels is large, and their sensitivities vary from one to another.
To visualize the data in a way that emphasizes the sensitivity to the exotic signals, we
follow Ref.~\cite{tevstudies}. 
Bins of the final discriminant for all channels are ordered by increasing signal-to-background ratio ($s/b$) and are shown
in comparison with predicted yields from signal and background processes
for the $J^P = 0^-$ and $J^P=2^+$ searches in Fig.~\ref{fig:lnsb} separately.
The backgrounds are fit to the data in each case, allowing the 
systematic uncertainties to vary within their a~priori constraints.
The exotic signals are normalized to the SM cross section times branching ratio multiplied by 
an exotic-signal scaling factor, $\mu_{\rm{exotic}}$. They are
shown in Fig.~\ref{fig:lnsb} 
with $\mu_{\rm{exotic}}=1$. The scaling factor for the SM Higgs boson signal is denoted by $\mu_{\rm{SM}}$. 
A value of one for either $\mu_{\rm{SM}}$ or $\mu_{\rm{exotic}}$ corresponds to a cross section times branching ratio as predicted for the SM Higgs boson. 
Both figures show agreement between the background predictions
and the observed data over five orders of magnitude with no evidence for an excess of exotic signal-like candidates.

We follow Ref.~\cite{tevstudies} and perform both Bayesian and modified frequentist
calculations of the upper limits on exotic $X$ boson production with and without SM Higgs production, best-fit cross sections allowing for the
simultaneous presence of a SM Higgs boson and an exotic $X$ boson, and hypothesis tests for 
signals assuming various production
rate times branching ratio values for the exotic bosons.
Both methods use likelihood calculations based on Poisson
probabilities that include SM background processes and 
signal predictions for the SM Higgs and exotic bosons multiplied by their respective scaling factors,
$\mu_{\rm{SM}}$ and $\mu_{\rm{exotic}}$. 
Systematic uncertainties on the predicted
rates and on the shapes of the distributions and their correlations are treated as described in Ref.~\cite{tevstudies}.
Theoretical uncertainties in cross sections and branching
ratios are considered fully correlated between CDF and D0, and between analysis samples. 
The uncertainties on the measurements of the integrated luminosities, which
are used to normalize the expected signal yields and the MC-based background rates, are 6.0\% (CDF)
and 6.1\% (D0).  Of these values, 4\% arises from 
the inelastic $p{\bar{p}}$ cross section~\cite{inelppbar}, which is fully correlated between CDF and D0.
The dominant uncertainties on the backgrounds are constrained by the data  
in low $s/b$ regions of the discriminant distributions.
Different methods were used by CDF and D0 to estimate $V$+jets and multijet 
backgrounds and so their uncertainties are considered uncorrelated.
Similarly, the uncertainties on the data-driven estimates of the $b$-tag 
efficiencies are considered uncorrelated between CDF and D0, as are the
uncertainties on the jet energy scales, the trigger efficiencies, and lepton identification efficiencies.
We quote Bayesian upper limits and best-fit cross sections assuming uniform priors for non-negative signal
cross sections,
and we use the modified frequentist method to perform the hypothesis tests.  
Systematic uncertainties are parameterized
by nuisance parameters with Gaussian priors, truncated so that no predicted yield for any process in any
search channel is negative.

\begin{figure}[t]
\begin{center}
\includegraphics[width=0.8\columnwidth]{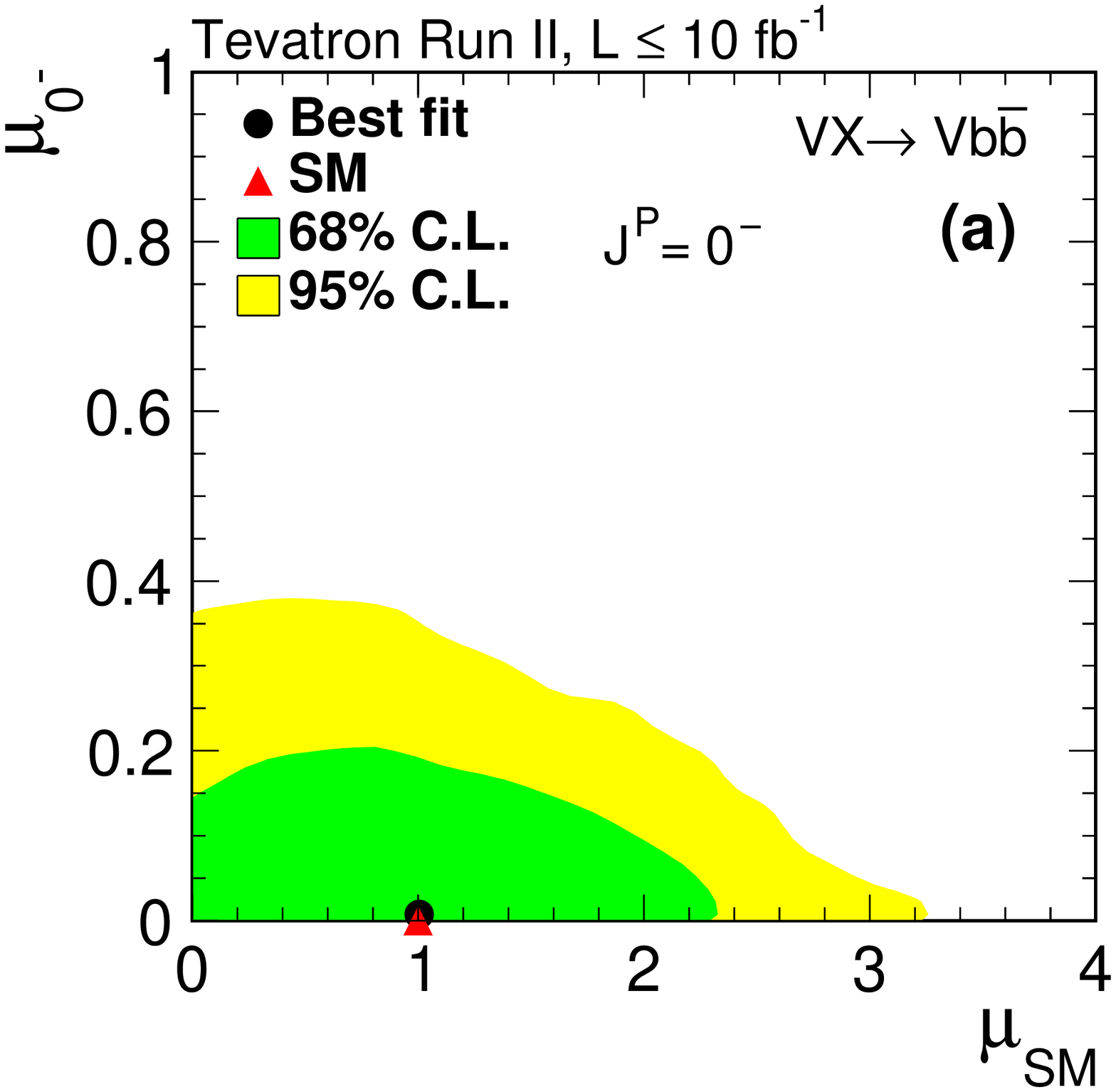}
\includegraphics[width=0.8\columnwidth]{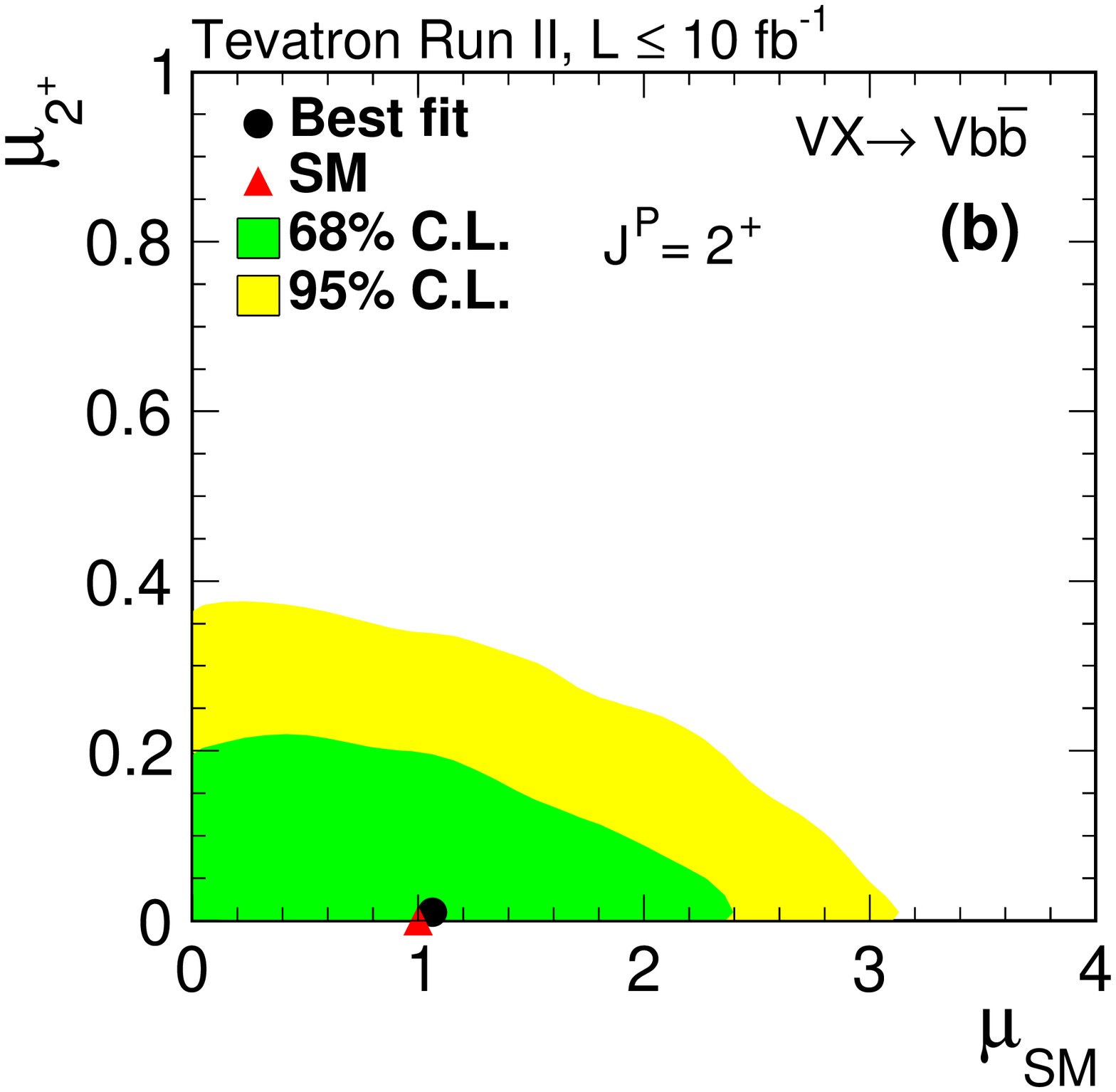}
 \caption{
 \label{fig:2dcomb} (color online).
Two-dimensional credibility regions in the 
($\mu_{\rm{exotic}}$, $\mu_{\rm{SM}}$) plane, for the combined CDF and D0 searches for 
(a) the pseudoscalar ($J^P=0^-$) boson, and
(b) the graviton-like ($J^P=2^+$) boson.
}
\end{center}
\end{figure}

\begin{figure}[t]
\includegraphics[width=0.99\columnwidth]{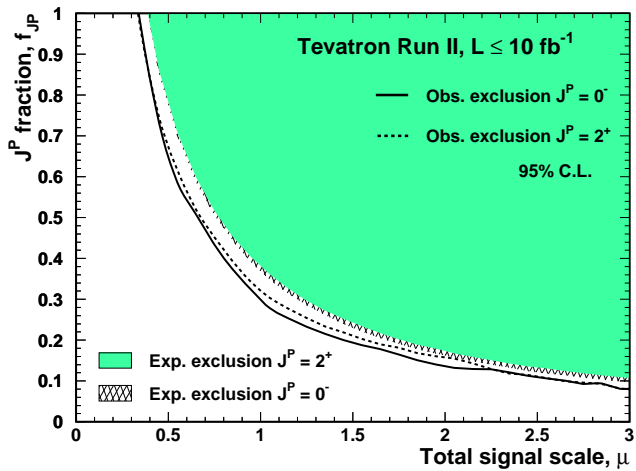}
\caption{ \label{fig:fraclimits} (color online).
Observed and expected upper limits at the 95\% C.L. on the fraction of exotic boson production for the $J^P=0^-$ and $J^P=2^+$ hypotheses.}
\end{figure}

For both the $J^P=0^-$ and $J^P=2^+$ models, we compute two 95\% credibility
upper limits on $\mu_{\rm{exotic}}$, one assuming $\mu_{\rm{SM}}=1$
and the other assuming $\mu_{\rm{SM}}=0$.  The expected limits are the 
median expectations assuming no exotic boson is present.
The results are listed in Table~\ref{tab:1dlimits}.  Two-dimensional credibility regions, which are the smallest
regions containing 68\% and 95\% of the posterior probabilities,
are shown in Fig.~\ref{fig:2dcomb}.
The points in the ($\mu_{\rm{SM}}$, $\mu_{\rm{exotic}}$) planes that
maximize the posterior probability densities are shown as the best-fit values.
These best-fit values are 
($\mu_{\rm{SM}}$=1.0, $\mu_{0^-}=0$) for the search for the $J^P=0^-$ state, and
($\mu_{\rm{SM}}$=1.1, $\mu_{2^+}=0$) for the search for the $J^P=2^+$ state.
We also derive upper limits on the fraction $f_{JP}=\mu_{\rm{exotic}}/(\mu_{\rm{exotic}}+\mu_{\rm{SM}})$, as functions of
the total $\mu=\mu_{\rm{exotic}}+\mu_{\rm{SM}}$, assuming a uniform prior probability density in non-negative $f_{JP}$, extended to
include fractions larger than 1.0 in order not to saturate the limits at $f_{JP}=0.95$ 
for $\mu<0.6$, where the test is weak.   The results are shown in Fig.~\ref{fig:fraclimits}.

In the modified frequentist approach~\cite{cls,pflh} we compute $p$~values 
for the discrete two-hypothesis tests, the SM Higgs 
boson hypothesis (the ``null'' hypothesis)
($\mu_{\rm{SM}}$=1, $\mu_{\rm{exotic}}$=0) and the exotic (``test'') hypothesis ($\mu_{\rm{SM}}$=0, $\mu_{\rm{exotic}}$=1),
both assuming that SM background processes are present.   The choice of setting $\mu_{\rm{exotic}}=1$ in the test hypothesis is
arbitrary; the sensitivity of the test is reduced if a smaller value is
assumed.  We use the log-likelihood ratio, LLR, 
defined to be $-2\ln (p({\rm{data}}|{\rm{test}})/p(\rm{data}|\rm{null})$,
where the numerator and denominator are maximized over systematic
uncertainty variations~\cite{tevstudies}.  The LLR distributions 
are shown in the supplemental material~\cite{epaps}.
We define the $p$~values $p_{\rm{null}} = P({\rm{LLR}\leq {\rm{LLR}}_{\rm{obs}}}|{\rm{SM}})$
and $p_{\rm{test}} = P({\rm{LLR}\geq {\rm{LLR}}_{\rm{obs}}}|{\rm{exotic}})$.  
The median expected $p$~values $p^{\rm exotic}_{\rm{null,med}}$ in the test hypothesis
and  $p^{\rm SM}_{\rm{test,med}}$ in the SM hypothesis quantify the sensitivities 
of the two-hypothesis tests for exclusion and discovery, respectively.
Table~\ref{tab:pvalues} lists these $p$~values for both exotic models, as well as
${\rm CL}_{\rm s}=p_{\rm{test}}/(1-p_{\rm{null}})$~\cite{cls} for the Tevatron combination.
To compute  $p_{\rm{test}}$ and the expected values of $p_{\rm{null}}$ and $p_{\rm{test}}$,
Wilks's theorem is used~\cite{aww}.

The similarity of the limits and $p$~values obtained for the
$J^P=0^-$ and the $J^P=2^+$ searches is expected since the exotic models predict 
excesses in similar portions of kinematic space.

\begin{table}
\begin{center}
\caption{\label{tab:1dlimits}
 Observed and median expected Bayesian upper limits at the 95\% credibility level on $\mu_{\rm{exotic}}$ for the
pseudoscalar ($J^P=0^-$) and
graviton-like ($J^P=2^+$) 
boson models, assuming either that the SM Higgs boson
is also present ($\mu_{\rm{SM}}=1$) or absent ($\mu_{\rm{SM}}=0$).}
\begin{ruledtabular}
\begin{tabular}{lcc}
Channel & Observed & Median Expected \\ 
  & (Limit/$\sigma_{SM}$) & (Limit/$\sigma_{SM}$) \\
 \hline
$J^P=0^-$, $\mu_{\rm{SM}}=0$   &     0.36 &        0.32 \\
$J^P=0^-$, $\mu_{\rm{SM}}=1$   &     0.29 &        0.32 \\
$J^P=2^+$, $\mu_{\rm{SM}}=0$   &     0.36 &        0.33 \\
$J^P=2^+$, $\mu_{\rm{SM}}=1$   &     0.31 &        0.34 \\
\end{tabular}
\end{ruledtabular}
\end{center}
\end{table}

\begin{table}
\begin{center}
\caption{\label{tab:pvalues}
Observed (obs) and median expected (med) LLR values and $p$~values for the combined
CDF and D0 searches for 
the pseudoscalar ($J^P=0^-$) boson and
the graviton-like ($J^P=2^+$) boson.
The $p$~values are listed, and the corresponding significances in 
units of standard deviations, using a one-sided Gaussian tail calculation, are given in parentheses.  The two hypotheses tested
are $(\mu_{\rm SM},\mu_{\rm{exotic}})=(1,0)$ and $(0,1)$ for the SM and the exotic models, respectively.
}
\vspace{0.2cm}
\begin{ruledtabular}
\begin{tabular}{clccc} 
&Analysis                                                   &  $J^P=0^-$                 &  $J^P = 2^+$               \\
\hline							                                
&\rule[-1mm]{0mm}{5mm}LLR$_{\rm{obs}}$                      &   27.1                     & 25.7                       \\
&\rule[-1mm]{0mm}{5mm}LLR$_{\rm med}^{\rm SM}$              &   23.7                     & 21.8                       \\
&\rule[-1mm]{0mm}{5mm}LLR$_{\rm med}^{\rm exotic}$          &   $-29.9$                  & $-29.6$                    \\
&\rule[-1mm]{0mm}{5mm}$p_{\rm{null}}$                       &  0.63 ($-0.34$)            & 0.66 ($-0.41$)             \\
&\rule[-1mm]{0mm}{5mm}$p_{\rm null, med}^{\rm exotic}$      &  1.8$\times 10^{-8}$ (5.5) & 1.9$\times 10^{-8}$ (5.5)  \\
&\rule[-1mm]{0mm}{5mm}$p_{\rm{test}}$                       &  9.4$\times 10^{-8}$ (5.2) & 1.9$\times 10^{-7}$ (5.1)  \\
&\rule[-1mm]{0mm}{5mm}$p_{\rm test, med}^{\rm SM}$          &  4.7$\times 10^{-7}$ (4.9) & 1.2$\times 10^{-6}$ (4.7)  \\
&\rule[-1mm]{0mm}{5mm}CL$_{\rm{s}}$                         &  2.6$\times 10^{-7}$ (5.0) & 5.6$\times 10^{-7}$ (4.9)  \\
&\rule[-1mm]{0mm}{5mm}CL$^{\rm SM}_{\rm{s, med}}$           &  9.4$\times 10^{-7}$ (4.8) & 2.3$\times 10^{-6}$ (4.6)  \\ 
\end{tabular}
\end{ruledtabular}
\end{center}
\end{table}

In summary, we combine CDF's and D0's tests for the presence of 
a pseudoscalar Higgs boson with $J^P=0^-$ 
and a graviton-like boson with $J^P=2^+$
in the $W\! X\rightarrow \ell\nu b{\bar{b}}$, the $ZX\rightarrow \ell^+\ell^- b{\bar{b}}$, and
the $V\! X\rightarrow\met b{\bar{b}}$ search channels using models described in Ref.~\cite{Ellis:2012xd}. 
The masses of the exotic bosons are assumed to be 125 GeV/$c^2$.
 No evidence is seen for either exotic particle, either in place of the
SM Higgs boson or produced in a mixture with a $J^P=0^+$ Higgs boson.  
In both searches, the best-fit cross section times the decay branching ratio
into a bottom-antibottom quark pair
of a $J^P=0^+$ signal component is consistent with the prediction of the SM Higgs boson. 
The Bayesian posterior probability densities for the $J^P=0^-$ and $J^P=2^+$ searches
are shown in Ref.~\cite{epaps}.

Upper limits at 95\% credibility on the rate of the production of an exotic 
Higgs boson in the absence of a SM $J^P=0^+$ signal are set at 0.36 times the SM Higgs 
production rate for both the $J^P=0^-$ and the $J^P=2^+$ hypotheses.
If the production rate of the hypothetical exotic particle times its branching ratio to
a bottom-antibottom quark pair is the
same as that predicted for the SM Higgs boson, then the exotic models are
excluded with significances of 5.0~s.d. and 4.9~s.d. for the $J^P=0^-$ and $J^P=2^+$
hypotheses, respectively.  

\begin{center}
{\bf Acknowledgments}  
\end{center}
We thank the Fermilab staff and technical staffs of the participating institutions for their vital contributions.  
We acknowledge support from 
the Department of Energy and 
the National Science Foundation (United States of America),
the Australian Research Council (Australia),
the National Council for the Development of Science and Technology and 
the Carlos Chagas Filho Foundation for the Support of Research in the State of Rio de Janeiro (Brazil),
the Natural Sciences and Engineering Research Council (Canada),
the China Academy of Sciences, the National Natural Science Foundation of China, and the National Science Council of the Republic of China (China),
the Administrative Department of Science, Technology and Innovation (Colombia),
the Ministry of Education, Youth and Sports (Czech Republic),
the Academy of Finland (Finland),
the Alternative Energies and Atomic Energy Commission and
the National Center for Scientific Research/National Institute of Nuclear and Particle Physics (France),
the Bundesministerium f\"{u}r Bildung und Forschung (Federal Ministry of Education and Research) and 
the Deutsche Forschungsgemeinschaft (German Research Foundation) (Germany),
the Department of Atomic Energy and Department of Science and Technology (India),
the Science Foundation Ireland (Ireland),
the National Institute for Nuclear Physics (Italy),
the Ministry of Education, Culture, Sports, Science and Technology (Japan),
the Korean World Class University Program and the National Research Foundation of Korea (Korea),
the National Council of Science and Technology (Mexico),
the Foundation for Fundamental Research on Matter (Netherlands),
the Ministry of Education and Science of the Russian Federation, 
the National Research Center ``Kurchatov Institute" of the Russian Federation, and 
the Russian Foundation for Basic Research (Russia),
the Slovak R\&D Agency (Slovakia), 
the Ministry of Science and Innovation, and the Consolider-Ingenio 2010 Program (Spain),
the Swedish Research Council (Sweden),
the Swiss National Science Foundation (Switzerland), 
the Ministry of Education and Science of Ukraine (Ukraine), 
the Science and Technology Facilities Council and The Royal Society (United Kingdom),
the A.P. Sloan Foundation (USA), 
and the European Union community Marie Curie Fellowship contract 302103.
%


\clearpage
\newpage

\begin{figure*}
\begin{center}
{\bf\LARGE  Supplemental Material}
\end{center}
\includegraphics[width=0.31\linewidth]{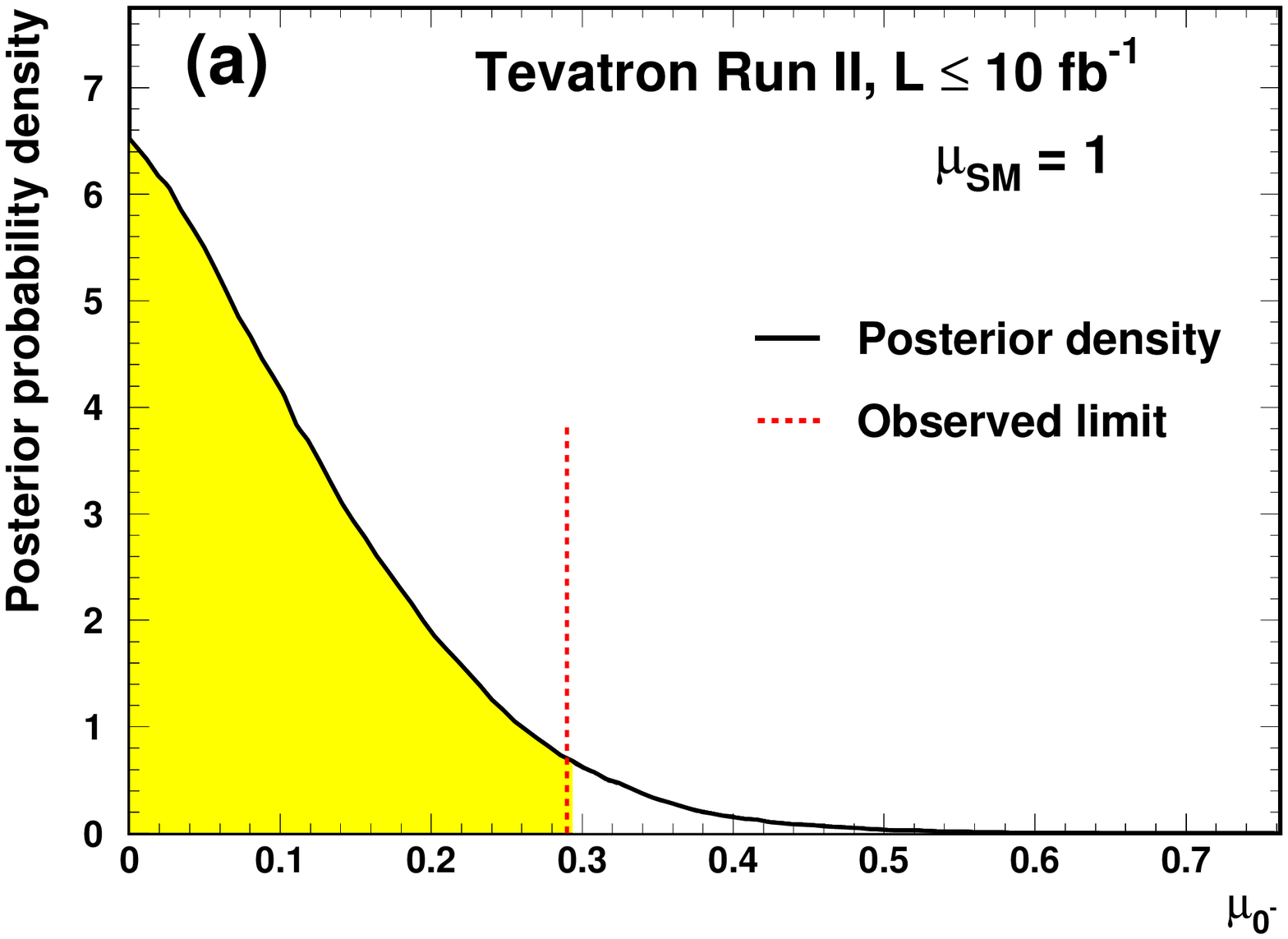}
\includegraphics[width=0.31\linewidth]{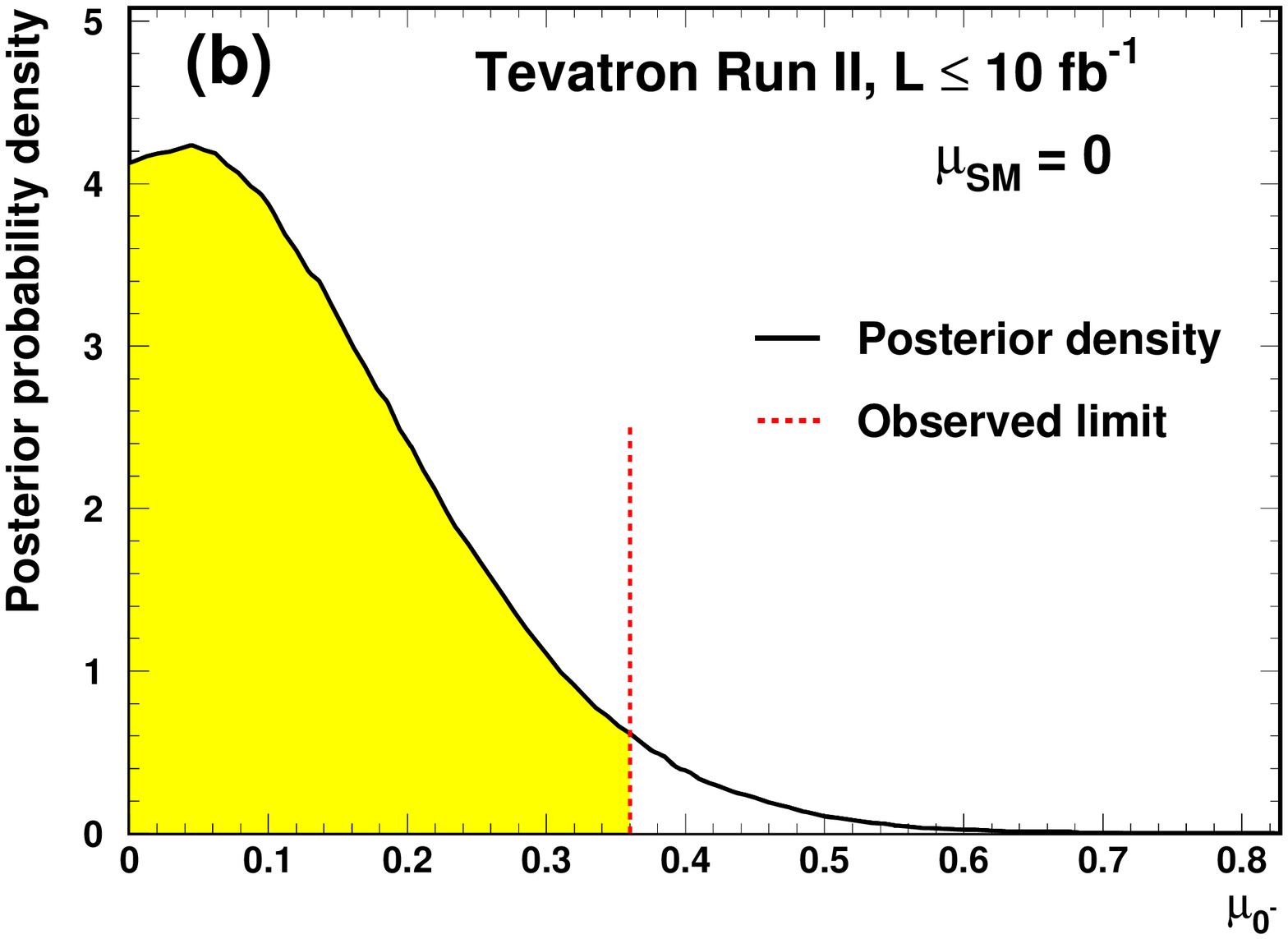}
\includegraphics[width=0.31\linewidth]{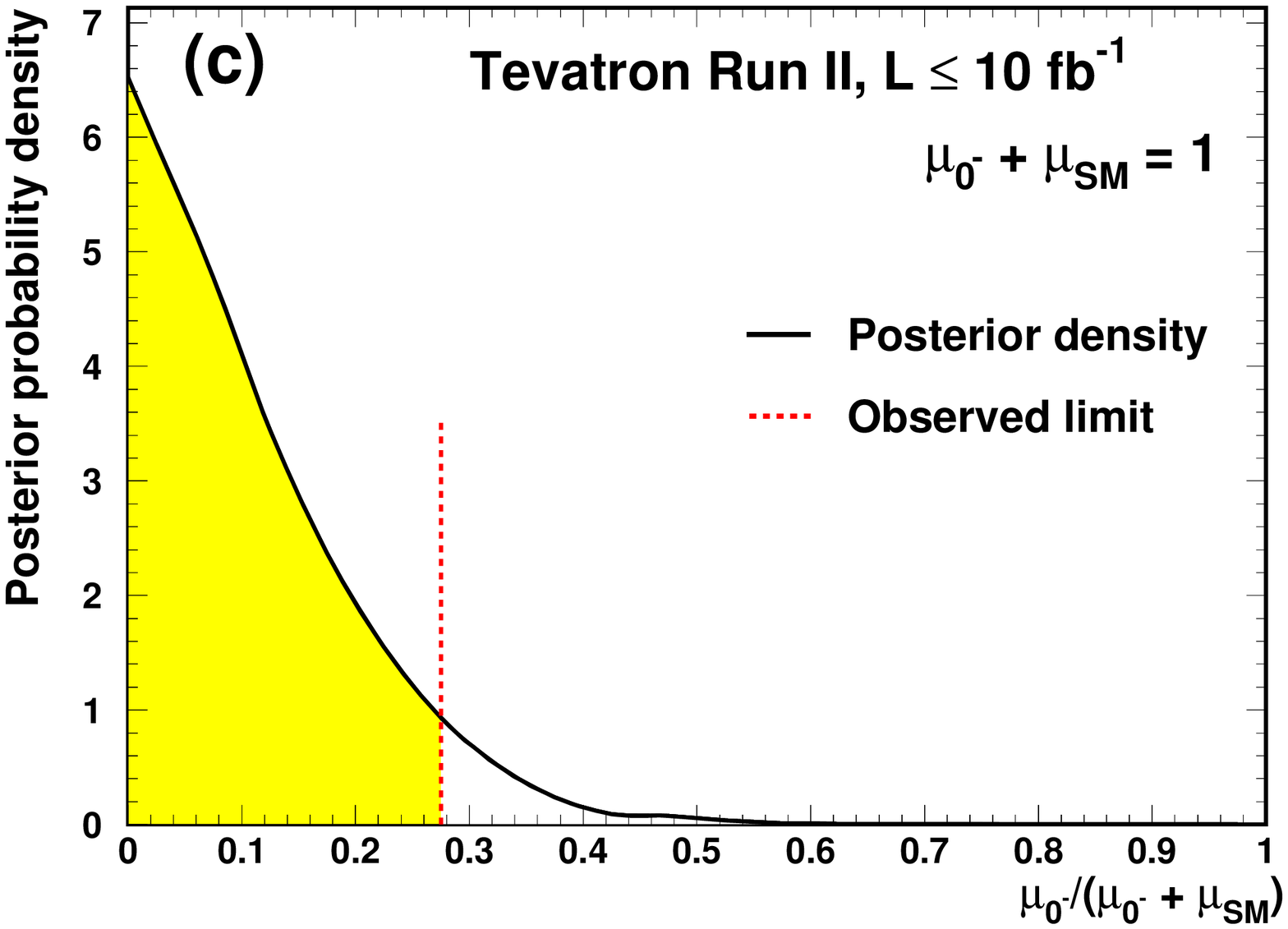}
\includegraphics[width=0.31\linewidth]{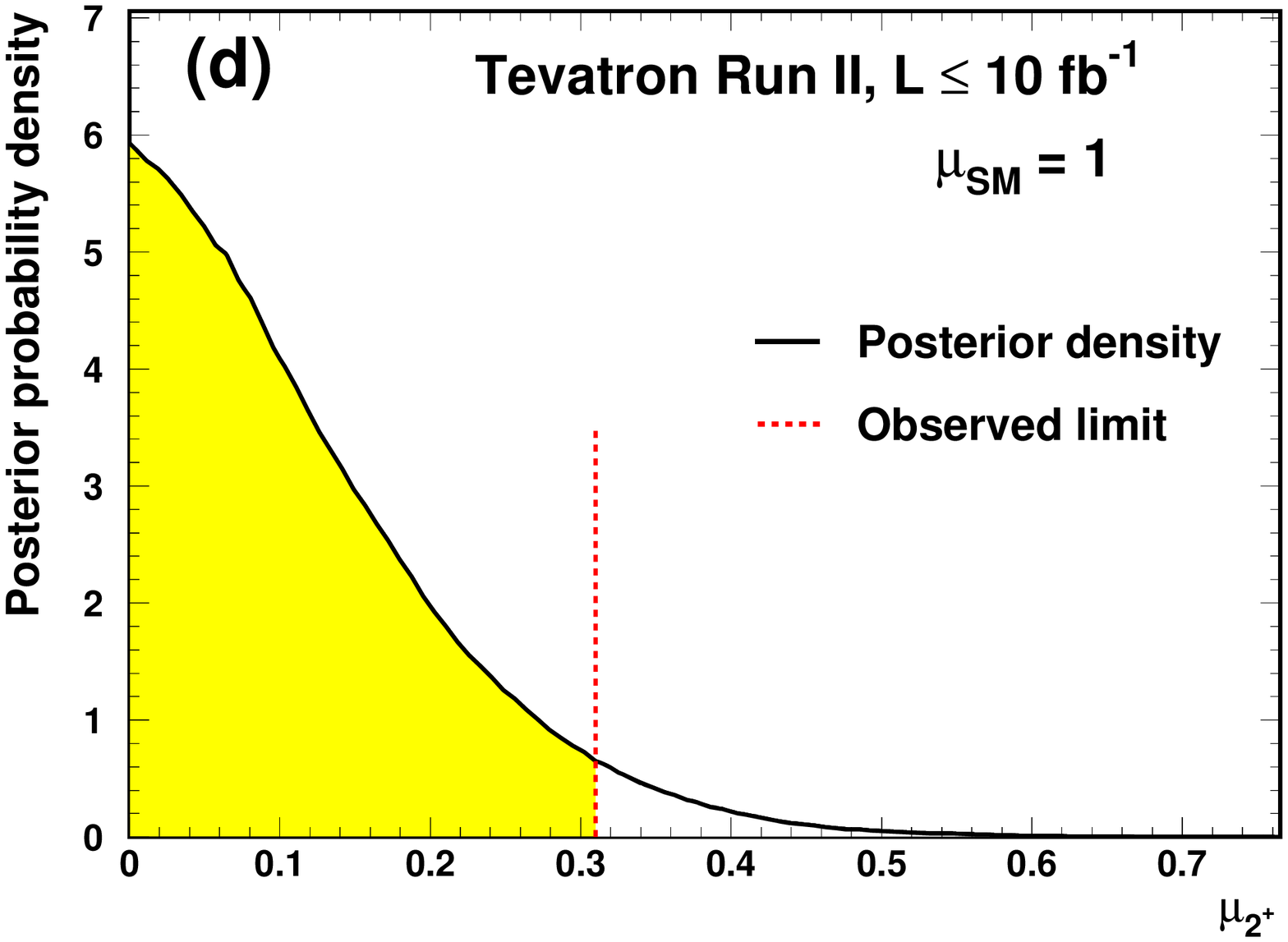}
\includegraphics[width=0.31\linewidth]{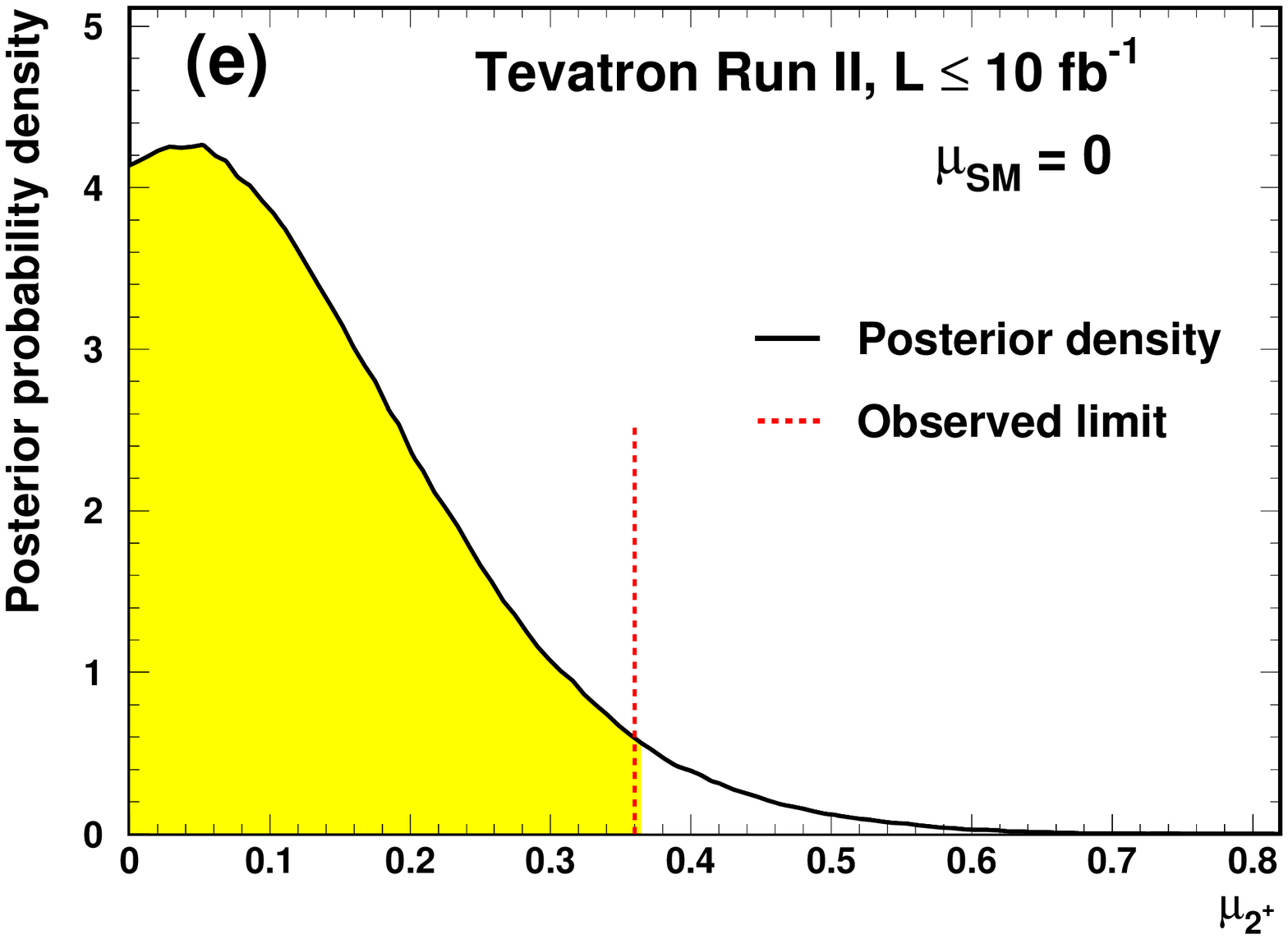}
\includegraphics[width=0.31\linewidth]{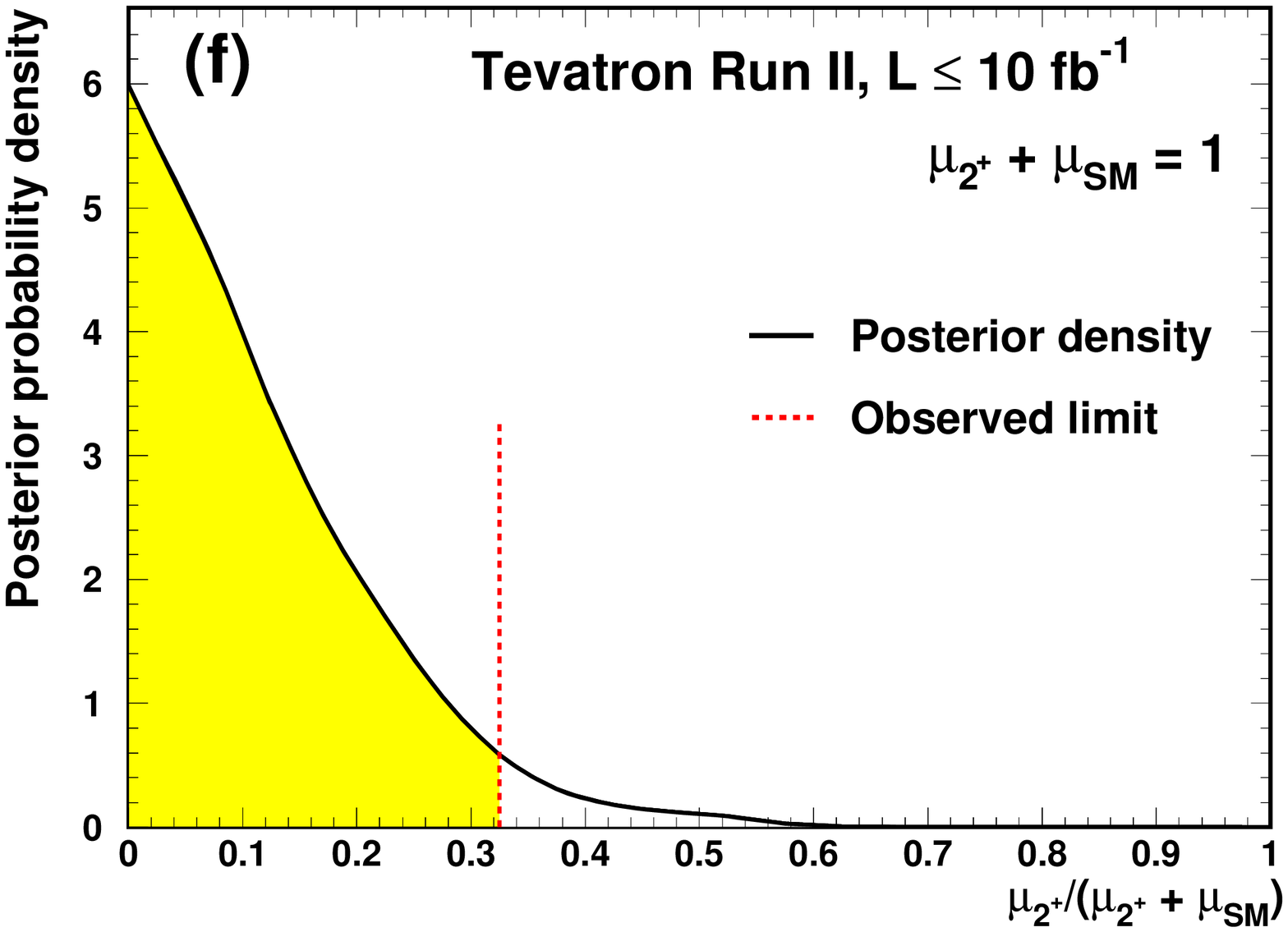}
\caption{ \label{fig:supplemental_pspost}  
Posterior probability density distributions for the combined searches for exotic $J^P=0^-$ and $J^P=2^+$ bosons.  
(a) The posterior probability density as a function of $\mu_{0^-}$ assuming $\mu_{\rm{SM}}=1$ and a uniform prior 
density for non-negative $\mu_{0^-}$, 
(b) the posterior probability density as a function of $\mu_{0^-}$ assuming $\mu_{\rm{SM}}=0$ and a uniform
density for non-negative $\mu_{0^-}$, 
and (c) the posterior probability
density as a function of the fraction of exotic boson production, $\mu_{0^-}/(\mu_{0^-}+\mu_{\rm{SM}})$, assuming 
$\mu_{0^-}+\mu_{\rm{SM}}=1$, and a uniform prior density for non-negative values of the fraction. The dashed vertical lines indicate the
observed limits.  Figures (d), (e), and (f) show the corresponding results for the $J^P=2^+$ boson searches.  }
\end{figure*}

\begin{figure*}
\begin{center}
\includegraphics[width=0.8\columnwidth]{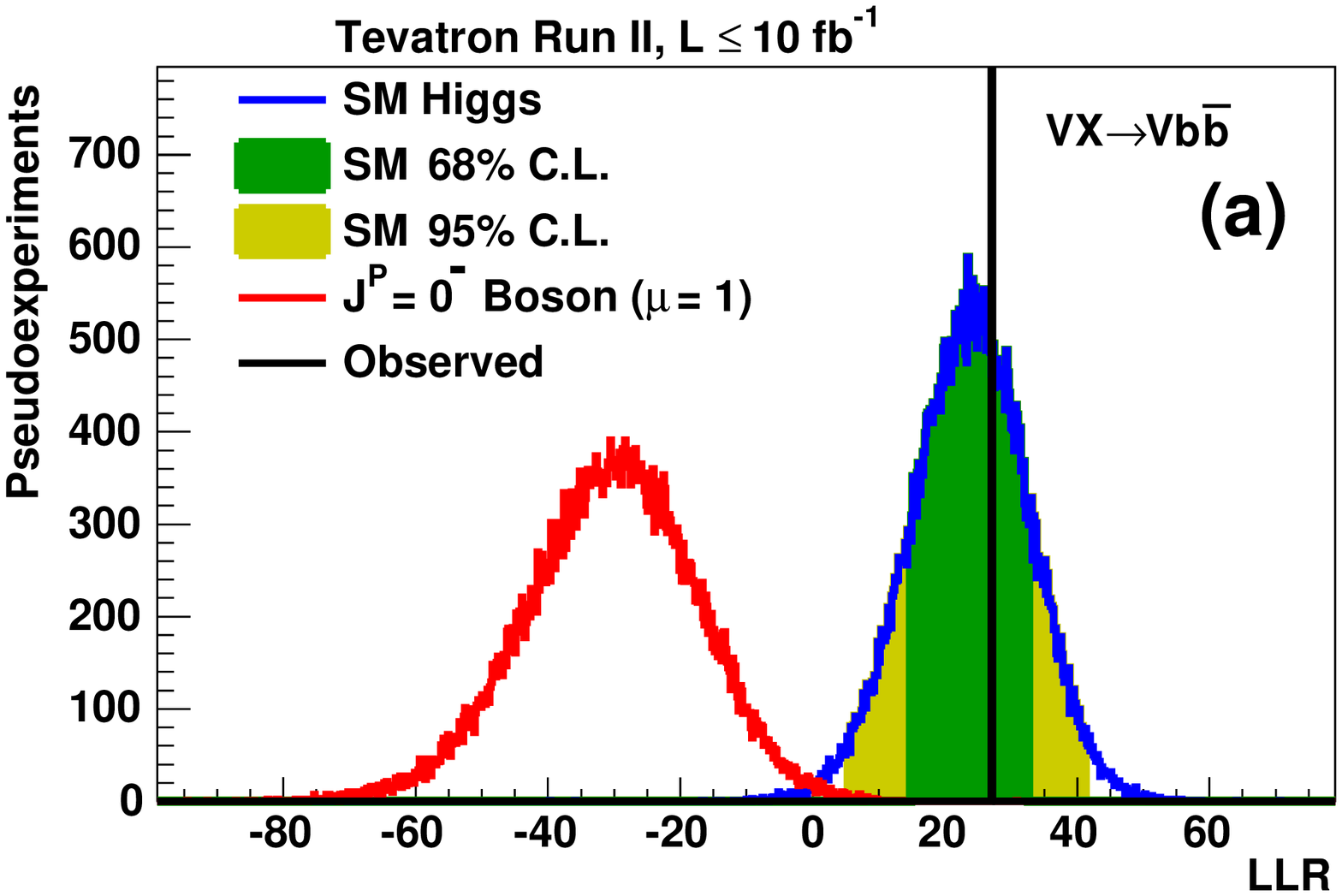}
\includegraphics[width=0.8\columnwidth]{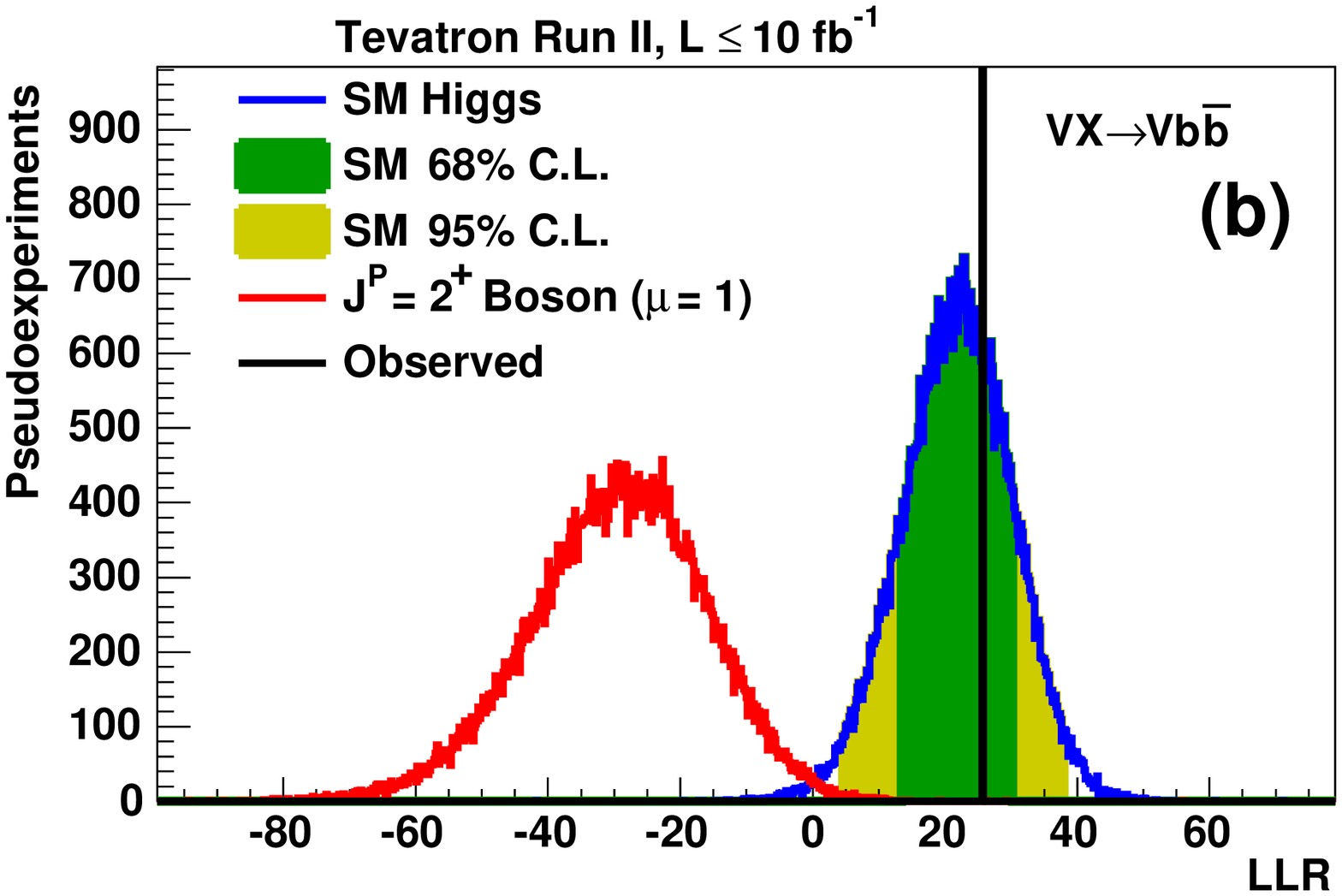}
 \caption{
 \label{fig:supplemental_llrcombmu1} 
Distributions of LLR for the combined CDF and D0 searches for 
(a) the pseudoscalar ($J^P=0^-$) boson, and
(b) the graviton-like ($J^P=2^+$) boson.
The LLR distributions are shown separately assuming that
an exotic particle is present with $\mu_{\rm{exotic}}=1$ plus SM backgrounds, 
and if the SM Higgs boson plus SM backgrounds are present.  The observed values of LLR are shown with vertical
lines.  Shaded regions show the 68\% and 95\% confidence level regions on the distributions assuming the SM Higgs boson
is present, centered on the median expectation.}
\end{center}
\end{figure*}


\begin{thebibliography}{000}
\bibitem{coord}
\bibinfo{note}{
 Positions and angles are
expressed in a cylindrical coordinate system, with the $z$~axis directed along the proton beam.
The polar angle $\theta$ is defined
with respect to the proton beam direction.
The missing transverse energy is defined as a sum over calorimeter towers,
$\vec{\not\!\!E}_{T} = -\sum_i  E_i\sin\theta_i {\hat n}_i$, where $i$ is the calorimeter tower number, and ${\hat n}_i$ is
a unit vector perpendicular to the beam axis and pointing to the $i$th tower.  The reconstructed $\vec{\not\!\!E}_{T}$ is corrected
for contributions from muons which register less energy in the calorimeter, and for jet energy corrections.  The scalar magnitude
of $\vec{\not\!\!E}_{T}$ is denoted ${\not\!\!E}_{T}$.
}
\bibitem[{mc()}]{mcbla}
\bibinfo{note}{In the CDF analyses, backgrounds from SM processes
with electroweak gauge bosons or top quarks are modeled using \PYTHIA~\cite{pythia}, 
\ALPGEN~\cite{alpgen}, \MCNLO~\cite{MC@NLO}, and \HERWIG~\cite{herwig}.  
For D0, these backgrounds are modeled using
\PYTHIA, \ALPGEN, and \SINGLETOP~\cite{comphep},
with \PYTHIA{} providing parton showering and hadronization for all the generators.}
\bibitem[{hfj()}]{hfjet}
\bibinfo{note}{A heavy-flavor jet is a reconstructed cluster of calorimeter
  energies associated with particles produced in the hadronization and decay of
  a $b$ or $c$ quark.}
\bibitem[{bta()}]{btagdef}
\bibinfo{note}{A $b$-tagged jet is a jet identified as being consistent with that
  expected from the decay products of a $b$ quark based on properties such
  as the presence of displaced track vertices or soft leptons.}
\end{thebibliography}
\end{document}